\documentstyle[twocolumn]{mn}

\input psfig

\def\beq{\begin{equation}}
\def\eeq{\end{equation}}
\def\bey{\begin{eqnarray}}
\def\eey{\end{eqnarray}}

\def\and{{\rm and\ }}
\def\be{\begin{equation}}
\def\ee{\end{equation}}

\def\spose#1{\hbox to 0pt{#1\hss}}
\def\lta{\mathrel{\spose{\lower 3pt\hbox{$\sim$}}
    \raise 2.0pt\hbox{$<$}}}
\def\gta{\mathrel{\spose{\lower 3pt\hbox{$\sim$}}
    \raise 2.0pt\hbox{$>$}}}

\input epsf

\title[Simulations of Centaurs I: Statistics]
{Simulations of the Population of Centaurs I: The Bulk Statistics}

\author[J. Horner, N.W. Evans \& M.E. Bailey]
       {J. Horner$^{1,2}$, N.W. Evans$^{2,3}$ \& M.E. Bailey$^4$\\
       $^1$ Physikalisches Institut, Universit\"at Bern, Sidlerstrasse
       5, CH-3012 Bern, Switzerland \\
       $^2$ Theoretical Physics, Department of Physics, 1 Keble Rd,
       Oxford OX1 3NP\\
       $^3$ Institute of Astronomy, Madingley Rd, Cambridge, CB3
       0HA\\
       $^4$ Armagh Observatory, College Hill, Armagh, BT61 9DG,
       Northern Ireland.}
\date{}
\pagerange{\pageref{firstpage}--\pageref{lastpage}}
\pubyear{}

\begin{document}
\maketitle
\label{firstpage}

\begin{abstract}
Large-scale simulations of the Centaur population are carried out.
The evolution of 23\,328 particles based on the orbits of 32 well-known
Centaurs is followed for up to 3Myr in the forward and backward
direction under the influence of the 4 massive planets. The objects
exhibit a rich variety of dynamical behaviour with half-lives ranging
from 540 kyr (1996 AR20) to 32 Myr (2000 FZ53). The mean half-life of
the entire sample of Centaurs is 2.7 Myr.  The data are analyzed
using a classification scheme based on the controlling planets at
perihelion and aphelion, previously given in Horner et al (2003).
Transfer probabilities are computed and show the main dynamical
pathways of the Centaur population. The total number of Centaurs with
diameters larger than 1 km is estimated as $\sim 44300$, assuming an
inward flux of one new short-period comet every 200 yrs. The flux into
the Centaur region from the Edgeworth-Kuiper belt is estimated to be 1
new object every 125 yrs. Finally, the flux from the Centaur region to
Earth-crossing orbits is 1 new Earth-crosser every 880 yrs.
\end{abstract}

\begin{keywords}
minor planets, asteroids -- planets and satellites: general --
celestial mechanics, stellar dynamics -- Kuiper belt
\end{keywords}

\section{Introduction}

The dynamical behaviour of the Centaurs is still poorly understood. It
is possible for a Centaur to work its way slowly inwards through the
outer Solar system, leading to eventual capture by Jupiter and
designation as a short-period comet. It is also possible for Centaurs
to drift outwards to join the Edgeworth-Kuiper belt, to be ejected
from the Solar system in an encounter with one of the massive, outer
planets, or even to be captured by these planets into temporary
satellite orbits. A small number may even impact upon the
planets. Therefore, the Centaurs potentially hold the key to
understanding the mechanisms by which the short-period comet
population is maintained, to explaining the distant, retrograde
satellites of the massive planets, and to allowing us a glimpse
of objects newly introduced to the Solar system from the
Edgeworth-Kuiper belt.

\begin{table}
\begin{center}
\begin{tabular}{ccc}\hline
 & & \\
Object & Perihelion & Aphelion \\ 
\null & (in au) & (in au) \\ 
 & & \\
J  & $4\lta q\lta6.6$ & $ Q\lta 6.6$ \\ 
JS & $4\lta q\lta6.6$ & $ 6.6\lta Q \lta 12.0$ \\ 
JU & $4\lta q\lta6.6$ & $ 12.0 \lta Q \lta 22.5$ \\ 
JN & $4\lta q\lta6.6$ & $ 22.5 \lta Q \lta 33.5$ \\ 
JE & $4\lta q\lta6.6$ & $ 33.5 \lta Q \lta 60.0$ \\ 
JT & $4\lta q\lta6.6$ & $ Q \gta 60.0$ \\ 
S      & $6.6 \lta q\lta 12.0$     & $Q \lta  12.0$  \\ 
SU     & $6.6\lta  q \lta 12.0$    & $12.0 \lta  Q \lta  22.5$  \\ 
SN     & $6.6\lta  q \lta 12.0$    & $22.5 \lta  Q \lta  33.5$  \\ 
SE     & $6.6\lta  q \lta 12.0$    & $33.5 \lta  Q \lta  60.0$  \\ 
ST     & $6.6\lta  q \lta 12.0$    & $Q \gta 60.0 $  \\ 
U      & $12.0\lta  q \lta 22.5$   & $Q \lta  22.5$  \\ 
UN     & $12.0\lta  q \lta  22.5$  & $22.5  \lta  Q \lta  33.5$  \\ 
UE     & $12.0\lta  q \lta  22.5$  & $33.5 \lta  Q \lta  60.0$  \\ 
UT     & $12.0\lta  q \lta  22.5$  & $Q \gta  60.0$  \\ 
N      & $22.5\lta  q \lta  33.5$  & $Q \lta  33.5$ \\ 
NE     & $22.5\lta  q \lta  33.5$  & $33.5 \lta  Q \lta  60.0$ \\ 
NT     & $22.5\lta  q \lta  33.5$  & $Q \gta  60.0$  \\ 
EK     & $33.5\lta  q \lta  60.0$  & $Q \lta  60.0$  \\ 
T      & $33.5\lta  q \lta  60.0$  & $Q \gta  60.0$  \\ 
 & & \\
\hline
 &  & \\
E  & $q \lta 4$ & $ Q \lta 4 $ \\ 
SP & $q \lta4$ & $ 4 \lta Q \lta 35 $ \\ 
I  & $q \lta4$ & $ 35 \lta Q \lta 1000$ \\ 
L  & $q \lta4$ & $ Q \gta 1000$ \\
 & & \\
\hline
\end{tabular}
\caption[] {The classification scheme introduced by Horner et
al. (2003). In the upper table, the first letter designates the planet
controlling the perihelion, the second letter the planet controlling
the aphelion or the region in which the aphelion lies, with the final
two classes EK and T being beyond all the giant planets. (J = Jupiter,
S = Saturn, U = Uranus, N = Neptune, EK = Edgeworth-Kuiper belt, T =
trans-Neptunian belt). The lower table refers to cometary bodies (E =
Encke, SP = short-period, I = intermediate and L = long-period).}
\label{tab:centaurs}
\end{center}
\end{table}

As early as 1990, when the only known Centaur was Chiron, it was
realised that such objects may lie on very unstable orbits.  Numerical
integrations by Hahn \& Bailey (1990) found that Chiron had a
half-life for ejection of around 1 Myr, but that the half-life to
become a short-period comet for the object was around only 200\,000
years, implying that Chiron could well have been a short-period comet
in the past and could possibly become one in the future. This is of
particular interest given the size of Chiron ($d \simeq 140-180$ km;
Groussin, Lamy \& Jorda 2004 and references therein) and other
Centaurs, because objects that large entering the inner Solar
system would be both spectacular and dangerous. In fact, the idea has
been mooted that objects of such size arrive in the inner Solar system
with some frequency, and then fragment, leading to swarms of debris
which have the potential to encounter the Earth.  The Kreutz
sun-grazer family may represent one example of such hierarchical
fragmentation, whereas other cases in which comets of more ordinary
size have undergone catastrophic fragmentation include 3D/Biela,
D/1994 (Shoemaker-Levy\,9) and C/1994\,S4 (LINEAR).  Whether such a
decay mode represents a {\sl generic\/} process in determining the
number of short-period comets can in principle be tested by examining
the differences in the size distribution of Jupiter-family comets from
those of their probable source objects, namely Centaurs,
Edgeworth-Kuiper belt objects, long-period comets and so on
(cf. Lowry, Fitzsimmons \& Collander-Brown 2003; Lamy et al. 2004).
Clube \& Napier (1984) have suggested that the Taurid meteoroid swarm
may be the relic of the last large object to undergo such a decay.

Despite the importance of the Centaurs, there has been little
systematic study of the population using numerical simulations.  Early
calculations on Chiron (Oikawa \& Everhart 1979, Hahn \& Bailey 1990)
and Pholus (Asher \& Steel 1993) identified the chaotic nature of
these two objects, though only small numbers of clones and modest
integration times ($<1$ Myr) were used. Work by Dones et al. (1996)
looked at the behaviour of four Centaurs (Chiron, Pholus, Nessus and
1994 TA) and two Jupiter-family comets (29P/Schwassmann-Wachmann 1 and
39P/Oterma). They found that the number of surviving objects decays
exponentially during the early part of the integrations, whilst the
decay becomes flatter after a number of life-times. They also noted
that Centaur half-lives inferred from numerical integrations are
smaller than those deduced from approximations like {\"O}pik's (1976)
theory and diffusion equations (e.g., van Woerkom 1948).  Levison \&
Duncan (1997) ran orbital integrations of 2200 test particles evolving
from the Edgeworth-Kuiper belt to short-period comets, passing through
the Centaur region in the process. The study of the integrations was
mainly focused on the behaviour of the objects both in the
Edgeworth-Kuiper belt and the cometary region, rather than in the
Centaur region itself. Hence, the dynamics of the Centaur population
remains largely unexplored.

In an earlier paper, Horner et al. (2003) introduced a new method of
classifying objects in the Solar system. This was based on the idea
that the dynamical evolution is largely determined by the planets that
control the perihelion and the aphelion of the object.  This
classification is particularly useful for the Centaurs, since it
breaks down the trans-Jovian region into 20 categories given in the
upper panel of Table~\ref{tab:centaurs}. Objects are labelled
according to the controlling giant planet, so, for example, a JS
object has perihelion under the control of Jupiter and aphelion under
the control of Saturn. Objects with perihelion distance $q \gta 33.5$
au are designated as either members of the Edgeworth-Kuiper belt (EK)
or trans-Neptunian disk (T).  In addition, objects with $q \lta 4$ au
are designated as comets and are subdivided into Encke types (E),
short-period (S), intermediate (I) and long-period (L), as summarised
in the lower panel of Table~\ref{tab:centaurs}.

The aim of this paper is to provide results from a detailed set of
simulations, exploiting the new classification scheme. The orbits of
32 of the best known Centaurs were used to create an ensemble of 23\,328
clones. The clones were integrated in the presence of the four massive
outer planets in both forward and backward directions for a period of
three million years, giving a vast data set with which to examine the
dynamics of their orbits. \S 2 describes the details of the numerical
simulations, whilst \S 3 provides half-lives for individual Centaurs.
\S 4 gives transition probabilities which allow the main dynamical
pathways through this region of the Solar system to be identified.
The simulations are used to estimate the total population of Centaurs
in \S 5, together with the typical fluxes inwards from the
Edgeworth-Kuiper belt. Finally, \S 6 considers possible correlations
between the dynamics and the colours of the Centaurs.

\begin{table*}
\begin{center}
\begin{tabular}{||c||c|c|c|c|c|c|c|c|c||}\hline \hline
Object & $a$ & $e$ & $i$ & $\omega$ & $\Omega$ & $H$ & $D_{\rm
min}$ & $D_{\rm max}$ & Class\\ \hline \hline
2000 GM137 &  7.853 & 0.118 & 15.9 & 123.5 &  89.7 & 14.3 & 4.7 & 13
&S  \\ \hline
1998 SG35  &  8.420 & 0.307 & 15.6 & 337.5 & 173.2 & 11.3 & 19& 52& JS
\\ \hline
2001 BL41  & 10.071 & 0.267 & 11.5 & 319.6 & 280.1 & 11.7 & 16& 43& SU
\\ \hline
2001 PT13  & 10.624 & 0.197 & 20.4 &  86.6 & 205.3 &  9.0 & 54& 150& SU
\\ \hline
2000 EC98  & 10.651 & 0.471 &  4.4 & 163.5 & 173.2 &  9.5 & 43& 120&JU
\\ \hline
1999 UG5   & 12.778 & 0.415 &  5.6 & 289.4 &  87.4 & 10.1 & 33& 90&SU \\ \hline
Chiron	   & 13.601 & 0.379 &  6.9 & 339.1 & 209.4 &  6.5 & 170& 470&SU \\ \hline
1996 AR20  & 15.197 & 0.627 &  6.2 & 107.9 & 330.1 & 14.0 & 5& 15&
JN \\ \hline
Chariklo   & 15.775 & 0.171 & 23.4 & 241.4 & 300.4 &  6.4 & 180& 490& U  \\ \hline
2001 XZ255 & 16.039 & 0.043 &  2.6 & 294.2 &  77.8 & 11.1 & 21& 57& U
\\ \hline
2000 QC243 & 16.560 & 0.203 & 20.7 & 150.0 & 337.9 &  7.6 & 100& 280& U
\\ \hline
1994 TA    & 16.849 & 0.301 &  5.4 & 154.9 & 137.7 & 11.5 & 17& 47& SU
\\ \hline
2001 SQ73  & 17.485 & 0.177 & 17.4 & 304.2 &  16.3 &  9.6 & 41& 110& U
\\ \hline
2000 CO104 & 17.497 & 0.256 &  4.0 & 339.2 & 346.8 & 10.0 & 34& 94&U
\\ \hline
1999 XX143 & 17.886 & 0.458 &  6.8 & 214.9 & 103.8 &  8.5 & 68& 190&SN \\ \hline
Asbolus    & 17.938 & 0.619 & 17.6 & 290.3 &   6.1 &  9.0 & 54& 150&SN
\\ \hline
2002 GO9   & 19.418 & 0.277 & 12.8 &  92.0 & 117.4 &  8.5 & 68& 190&
UN \\ \hline
1998 QM107 & 20.042 & 0.136 &  9.4 & 154.9 & 127.2 & 10.4 & 28& 78& UN
\\ \hline
Pholus     & 20.265 & 0.573 & 24.7 & 354.6 & 119.3 &  7.0 & 140& 370&
SN \\ \hline
2002 CA249 & 20.713 & 0.385 &  6.4 & 182.4 & 313.6 & 12.0 & 14& 27& UN
\\ \hline
1999 HD12  & 21.322 & 0.583 & 10.1 & 288.8 & 177.7 & 12.8 & 9.4& 26&
SE \\ \hline
2002 DH5   & 22.433 & 0.384 & 22.5 & 323.7 & 157.0 & 10.4 & 28& 78& UN
\\ \hline
2002 GZ32  & 23.081 & 0.216 & 15.0 & 154.4 & 107.2 &  6.9 & 140& 390& UN \\ \hline
1995 SN55  & 23.564 & 0.663 &  5.0 &  49.3 & 144.6 &  6.0 & 220&
590 &SE \\ \hline
2000 FZ53  & 23.765 & 0.479 & 34.9 & 290.8 & 202.4 & 11.4 & 18& 49& UE \\ \hline
Nessus     & 24.404 & 0.517 & 15.7 & 170.1 &  31.4 &  9.6 & 41& 110& SE
\\ \hline
Hylonome   & 24.909 & 0.243 &  4.2 &   5.5 & 178.2 &  8.0 & 86& 240&
UN \\ \hline
2002 GB10  & 25.139 & 0.396 & 13.3 & 238.9 & 315.5 &  7.8 & 95& 260& UE \\ \hline
2001 KF77  & 25.992 & 0.240 &  4.4 & 266.4 &  14.6 &  9.4 & 45& 120&UN
\\ \hline
1998 TF35  & 26.429 & 0.383 & 12.6 & 301.8 &  52.0 &  9.3 & 47& 130&UE \\ \hline
2002 FY36  & 28.969 & 0.114 &  5.4 & 194.1 & 332.8 &  8.4 & 72& 200&N  \\ \hline
2000 QB243 & 28.953 & 0.381 &  6.5 & 339.4 & 331.1 &  8.2 & 79& 220&UE \\ \hline
\hline
\end{tabular}
\caption[The objects simulated] {\label{tab:orbits} The names of the
objects simulated, arranged in order of increasing semi-major axis,
together with their orbital elements as of June 2002. Here, $a$ is the
semi-major axis measured in astronomical units (au), $e$ is the
eccentricity, $i$ is the inclination (in degrees), $\omega$ is the
argument of perihelion (in degrees), $\Omega$ is the longitude of the
ascending node of the orbit (in degrees) and $H$ is the absolute
visual magnitude. $D_{\min}$ and $D_{\rm max}$ are the values for the
diameter (in km) assuming albedos of 0.15 and 0.02 respectively, which
are typical upper and lower limits from albedo measurements of comets
and Centaurs performed to date. `Class' is the classification of the
object, using the scheme given in Horner et al. (2003). [The data are
compiled from the Minor Planet Center.]}
\end{center}
\end{table*}

\section{Integrations}

In order to study the bulk statistics of Centaurs, 32 objects were
selected from the list of Centaurs given on the Minor Planet Center's
website\footnote{http://cfa-www.harvard.edu/iau/lists/Centaurs.html}. The
objects were restricted to those with an observed arc of at least 30
days and an aphelion distance of less than 40 au. This ensures that
only Centaurs with moderately well-determined orbits were included in
our sample.  The list of objects is given in Table \ref{tab:orbits}.
Over time, as the Centaurs are observed over longer arcs, the accuracy
with which the orbits are known increases, and the orbits given on the
Minor Planet website change accordingly.  The orbits used in these
integrations therefore represent the best available information as of
June 2002. Table~\ref{tab:orbits} also gives each Centaur's absolute
magnitude $H$, which is defined as the apparent magnitude the object
would have, if it were placed at both 1 au from the Earth and 1 au
from the Sun and was observed at zero phase angle. This is calculated
in ignorance of any out-gassing that might occur.  We can estimate the
maximum and minimum diameter, assuming values of the albedo between
0.15 and 0.02. This gives a crude reckoning of the size, though
photometric work is required to obtain any more detail. Of the objects
studied in these integrations, the one with the brightest absolute
magnitude is 1995 SN55, with a value of $H = 6.0$, which corresponds
to a diameter of between 220 and 590 km. The object with the faintest
absolute magnitude is 2000 GM137, with $H = 14.3$ giving a diameter of
between 5 and 13 km. This is similar to the size determined for some
cometary nuclei (e.g. Sanzovo et al. 2001; Lamy et al. 2004). Hence,
the Centaurs come in a wide range of sizes, from very large (1995 SN55
and Chiron) to those comparable in size with normal comets (2000
GM137).

The orbital elements of each object were used to create a swarm of 729
clones, distributed through a small cube of a-e-i (semimajor axis,
eccentricity and inclination) space, centred on the original orbit.
The clones of the objects were created by incrementally increasing
(and decreasing) the semi-major axis of the object by 0.005 au, the
eccentricity by 0.005, and the inclination by $0.01^\circ$. These
increments are sufficiently small that the clones can be considered as
initially essentially indentical to one another, yet they are large
enough to ensure that the subsequent chaotic dynamical evolution
following close planetary encounters rapidly disperse their orbits
through phass space.  So, nine values were used for each of these
elements, with the central (fifth) value of the nine having the
original orbital elements. This gives 729 clones of each of the 32
Centaurs, giving a grand total of 23\,328 objects.

The use of multiple clones of an individual object in the study of its
behaviour over time is desirable for a number of reasons. First, the
observations contain some uncertainty, which means that the orbit
itself is not known beyond a certain degree of precision. This alone
would be enough to promote the use of a cluster of orbits with
slightly different parameters. In addition, the chaotic nature of the
orbits implies that an infinitesimally small change in the initial
parameters may lead to a major difference in the final outcome of the
simulation. This means that, beyond a certain time in the future, an
object could be anywhere within the Solar system, or even beyond, as a
result of a tiny change in the initial elements. These two facts taken
together suggest that the best means to examine the future or past
behaviour of an object is to integrate a large number of clones, and
to examine the statistical properties of the dataset (Hahn \& Bailey
1990, Dones et al. 1996). The number of clones used in such a
simulation is chosen to maximise the size of the dataset available for
analysis, without requiring an excessive amount of time for the
simulations to run. The simulations described here took about three
months to run on a desktop workstation.

The clusters of Centaurs were then integrated for 3 Myr in both the
forward and backward directions. The gravitational influence of the
four Jovian planets (Jupiter, Saturn, Uranus and Neptune) was included
in the integrations, which were all carried out using the hybrid
integrator within the {\sc Mercury} (Chambers 1999) software
package. This is a symplectic integrator, which makes use of a
turnover function to switch to an accurate Bulirsch-Stoer algorithm
for close encounters.  The terrestrial planets (Mercury, Venus, Earth
and Mars) were all omitted from the integrations, and their masses
added to that of the Sun.  The only slight detriment that this causes
is loss of accuracy when objects are captured into orbits crossing
those of the terrestrial planets. Even in this case, however, the
effects of Jupiter (and the other giant planets, if the object's
aphelion lies sufficiently far from the Sun) are generally much
greater than those of the terrestrial planets.

After some trials, a time step of 120 days was used for the
integrations, since this was found to give a good compromise between
speed and accuracy.  In order to determine the most efficient
timestep, an object was placed on a typical short period cometary
orbit with perihelion near the Earth and aphelion near Jupiter. A
number of clones were created, and the ensemble was integrated for
$10^5$ yrs with time steps of 30, 60, 120, 240 and 360 days. The
resulting orbital elements were then compared, and it was found that
the results for timesteps of 30, 60, 120 and 240 days gave consistent
results, while 360 days was too long a time step. After a number of
such trials, a time step of 120 days was used for the integrations,
since this was found to give a good compromise between speed and
accuracy, the mid-range value of 120 days being chosen so as to err on
the side of accuracy where possible.

An ejection distance of 1000 au from the Sun was used, following
Levison \& Duncan (1994).  Any object which reached this distance was
removed from the integration~\footnote{The long-period comets (the L
class in Table~\ref{tab:centaurs}) are defined to have aphelion $Q$ in
excess of 1000 au. Our choice of ejection distance means that the
statistical properties of the L class cannot be reliably computed from
our simulations.}. Also removed were those objects which impacted upon
the surface of the Sun ($q < 0.005$ au), or on any one of the giant
planets (the separation is less than the physical radius of the
planet). On completion of the integration, datafiles were created for
each clone which gave the values of the orbital elements at 100 year
intervals for the clone's entire lifetime within the integration. It
was on these files that the analysis was carried out. Each one of
these files was approximately 5 Mbytes in size (the exact size varied,
since the file terminated with the ejection of the clone from the
simulation). Hence, the 23\,328 clones in total occupied $\approx 120$
Gbytes of disk space for their orbital elemental evolution alone,
prior to any analysis -- a daunting dataset by any standards!

\section{Half-Lives}

\begin{figure*}
\centerline{\epsfxsize=10cm \epsfbox{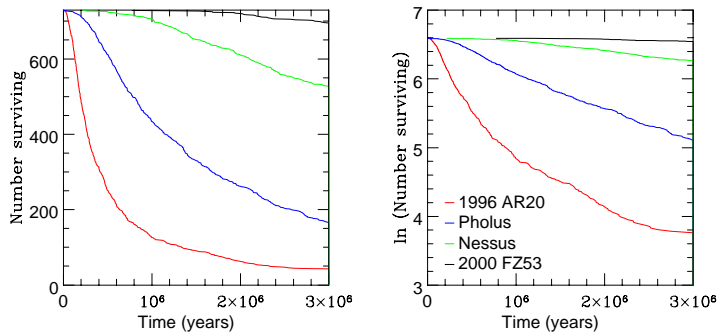}}
\caption{The number of surviving clones (lower left panel) and its
natural logarithm (lower right panel) as a function of time for 1996
AR20, Pholus, Nessus and 2000 FZ53. The upper panels show the same
quantities but now for the entire ensemble of 23\,328 clones. In all
cases, the graphs are deduced from the simulations in the forward
direction.}
\label{fig:DecayGraph}
\end{figure*}

It is straightforward to calculate the value of the half-life for each
Centaur.  As illustrated in Figure \ref{fig:DecayGraph}, the number of
clones remaining within the simulation decays in a roughly exponential
manner as a function of time. The four objects whose decay is shown
are 1996 AR20 (the object with the shortest lifetime of those
studied), Pholus (an object with a moderately short lifetime), Nessus
(a relatively long lived object) and 2000 FZ53 (the object with the
longest lifetime).  It has often been noted in long Solar system
integrations that the number of clones remaining within a simulation
decays exponentially with time (e.g., Dones et al. 1996, Holman 1997,
Evans \& Tabachnik 1999). The trait is less obvious with the longer
lived objects. For example, Dones et al. (1996) found that the number
of surviving clones in their integrations decayed exponentially at
early times, while at later times (generally greater than twice the
half-life) the decay was slower. This is because those objects
surviving the longest are the ones transferred to the most stable
areas of the Solar system.

The number of clones $N$ which remain after time $t$ is therefore given
by
\begin{equation}
N = N_0  e^{ - \lambda t },\qquad\qquad
\lambda = {{0.693}\over T_{1 \over 2}},
\label{eq:half-life}
\end{equation}
where $N_0$ is the initial number of clones, $\lambda$ is the decay
constant, and $T_{1 \over 2}$ is the half-life for the object.  To
calculate the half-life, the simulation data were analysed with the
help of least-square fitting routines from Press et al. (1992). The
software provides the value of the $\chi^2$ function
\begin{equation}
 \chi^2(\lambda) = \sum_{i=1}^N \left[{{\ln N_i - \ln N_0 + \lambda
 t_i} \over \sigma _i}\right]^2
\end{equation}
where $N_i$ is the number of clones remaining at time $t_i$ and the
$\sigma _i$ are the individual standard deviations on the data points.
As the $\sigma_i$ are unknown, we proceed by first assigning uniform
errors, fitting for the model parameters by minimizing the $\chi^2$
and then rescaling the errors using eqn. [15.1.6] of Press et
al. (1992). Of course, this well-known procedure precludes an
independent estimate of the goodness of fit.

The overall dataset of 23\,328 objects has an ensemble half-life of
2.76 Myr in the forward direction and 2.73 Myr in the backward
direction. This gives us an estimate of the mean lifetime of a typical
Centaur.  Note that this lifetime adds weight to the argument that the
population of the Centaur region is in a steady state (a reservoir of
objects constantly being drained by Jupiter, and refilled from a
long-lived source like the Edgeworth-Kuiper belt).

Dones et al. (1996) calculated the half-lives of Chiron, Pholus,
Nessus and 1994 TA using $\sim 100$ clones. Their results for Chiron
and Nessus are in excellent agreement with ours, but they found $T_{1
\over 2}$ = 2.1 and 2.4 Myr for Pholus and 1994 TA -- somewhat larger
than our results. The most likely cause of the discrepancy is in the
different algorithms used to populate the clones. Dones et al. carried
out a comparison with approximations like {\"O}pik's (1976) theory and
diffusion equations (e.g., van Woerkom 1948) and concluded that both
methods significantly overestimate the lifetimes by factors between 2
and 5. It seems that numerical simulations with large numbers of
clones are the only reasonably reliable method for half-life
estimation.

\begin{table*}
\begin{tabular}{||c|c|c|c|c| |c|c|c|c|c||}
 \hline \hline
 Object      & C  & D & $T_{1 \over 2}$ & $\sigma$   & Object  & C & D &  $T_{1 \over 2}$  & $\sigma$  \\ \hline \hline
 1996 AR20   & JN & F &  0.54  &  0.02 &  2001 SQ73  & U  & F &  2.86  &  0.11 \\ 
 1996 AR20   & JN & B &  0.59  &  0.02 &  2001 SQ73  & U  & B &  2.73  &  0.10\\ \hline
 2000 EC98   & JU & F &  0.61  &  0.02 &  2002 GO9   & UN & F &  2.93  &  0.11 \\ 
 2000 EC98   & JU & B &  0.63  &  0.02 &  2002 GO9   & UN & B &  3.67  &  0.14 \\ \hline
 1998 SG35   & JS & F &  0.67  &  0.03 &  2000 QC243 & U  & F &  3.18  &  0.12 \\ 
 1998 SG35   & JS & B &  0.65  &  0.02 &  2000 QC243 & U  & B &  3.44  &  0.13 \\ \hline
 2000 GM137  & S  & F &  0.72  &  0.03 &  2002 CA249 & UN & F &  4.06  &  0.15 \\ 
 2000 GM137  & S  & B &  0.68  &  0.03 &  2002 CA249 & UN & B &  2.54  &  0.09 \\ \hline
 1995 SN55   & SE & F &  0.70  &  0.03 &  1998 QM107 & UN & F &  4.87  &  0.18 \\ 
 1995 SN55   & SE & B &  0.80  &  0.03 &  1998 QM107 & UN & B &  5.65  &  0.21 \\ \hline
 1999 UG5    & SU & F &  0.74  &  0.03 &  Nessus     & SE & F &  4.91  &  0.18 \\ 
 1999 UG5    & SU & B &  0.85  &  0.03 &  Nessus     & SE & B &  6.40  &  0.24 \\ \hline
 Asbolus     & SN & F &  0.86  &  0.03 &  Hylonome   & UN & F &  6.37  &  0.24 \\ 
 Asbolus     & SN & B &  0.75  &  0.03 &  Hylonome   & UN & B &  7.30  &  0.27 \\ \hline
 2001 PT13   & SU & F &  0.94  &  0.04 &  2001 KF77  & UN & F &  8.89  &  0.33 \\ 
 2001 PT13   & SU & B &  0.87  &  0.03 &  2001 KF77  & UN & B &  10.1 &  0.4 \\ \hline
 2001 BL41   & SU & F &  0.95  &  0.04 &  2002 DH5   & UN & F &  9.08  &  0.34 \\ 
 2001 BL41   & SU & B &  0.95  &  0.04 &  2002 DH5   & UN & B &  12.8 &  0.5 \\ \hline
 Chiron      & SU & F &  1.03 &  0.04 &  2002 GZ32  & UN & F &  11.3 &  0.4    \\ 
 Chiron      & SU & B &  1.07 &  0.04 &  2002 GZ32  & UN & B &  7.78 &  0.28  \\ \hline
 1999 XX143  & SN & F &  1.06 &  0.04 &  Chariklo   & U  & F &  10.3 &  0.4 \\ 
 1999 XX143  & SN & B &  1.38 &  0.05 &  Chariklo   & U  & B &  9.38 &  0.35 \\ \hline
 1999 HD12   & SE & F &  1.22 &  0.05 &  1998 TF35  & UE & F &  11.5 &  0.4 \\ 
 1999 HD12   & SE & B &  1.13 &  0.04 &  1998 TF35  & UE & B &  10.8 &  0.4 \\ \hline
 Pholus      & SN & F &  1.28 &  0.05 &  2002 GB10  & UE & F &  11.1 &  0.4 \\ 
 Pholus      & SN & B &  1.39 &  0.05 &  2002 GB10  & UE & B &  13.1 &  0.5 \\ \hline
 1994 TA     & SU & F &  1.78 &  0.07 &  2002 FY36  & N  & F &  12.5 &  0.5 \\ 
 1994 TA     & SU & B &  1.52 &  0.06 &  2002 FY36  & N  & B &  13.5 &  0.5 \\ \hline
 2000 CO104  & U  & F &  1.89 &  0.07 &  2000 QB243 & UE & F &  13.0 &  0.5 \\ 
 2000 CO104  & U  & B &  2.24 &  0.08 &  2000 QB243 & UE & B &  17.8 &  0.7 \\ \hline
 2001 XZ255  & U  & F &  2.94 &  0.11 &  2000 FZ53  & UE & F &  26.8 &  1.0 \\ 
 2001 XZ255  & U  & B &  2.43 &  0.09 &  2000 FZ53  & UE & B &  32.3 & 1.2 \\ \hline \hline
\end{tabular} 
\caption[Half-Lives of the simulated objects] { \label{tab:half-lives}
Half-Lives $T_{1 \over 2}$ (in Myr) of the simulated objects, together
with an error estimate $\sigma$ (in Myr). The column labelled C gives
the class of the Centaur, while the column labelled D gives the
direction of integration (F for forward, B for backward).}
\vspace{0.5cm} 
\end{table*}
\begin{figure*}
\centerline{\epsfysize = 11cm \epsfbox{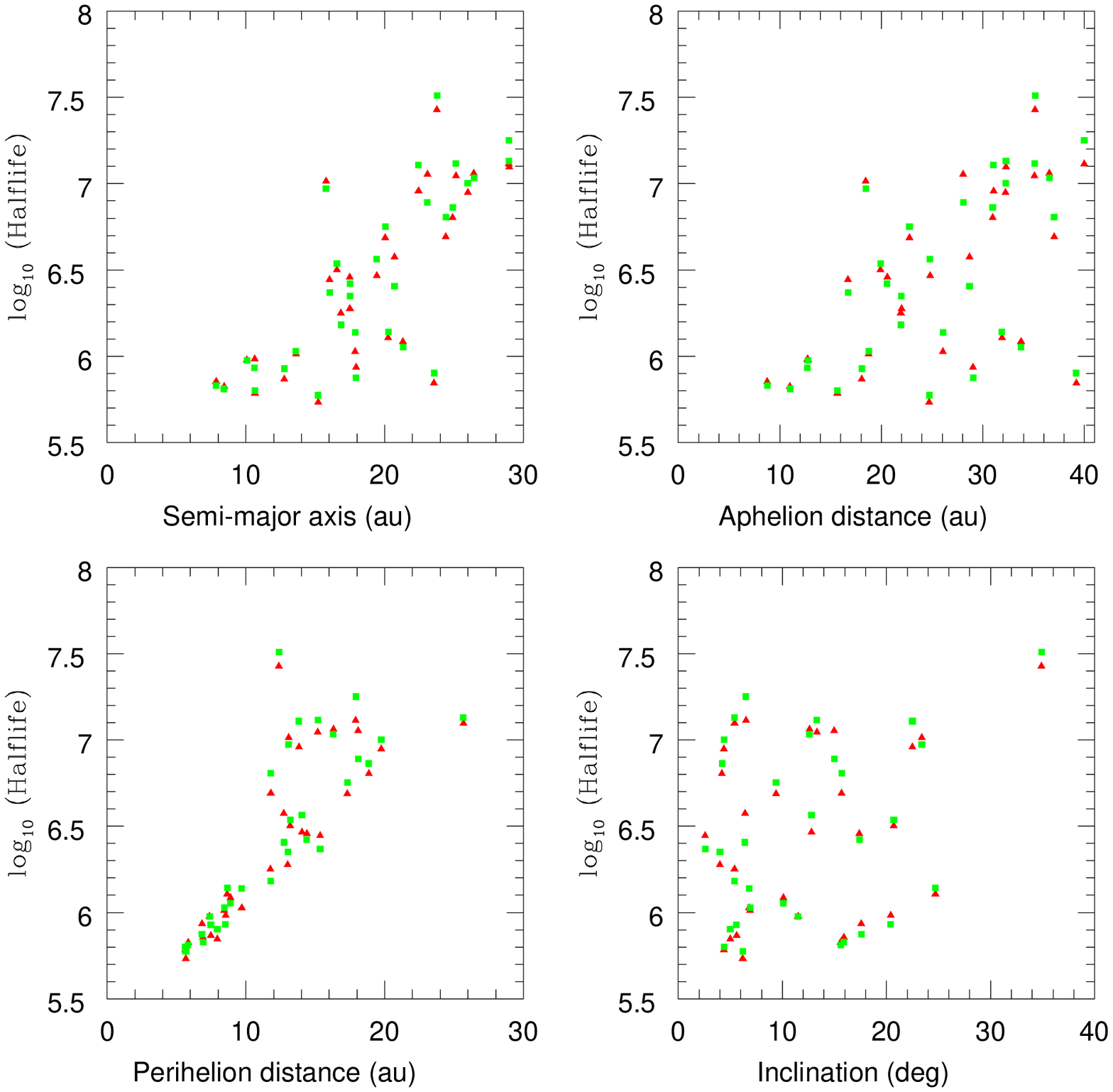}}
\caption[Natural logarithm of half-lives against orbital elements] {The
relationship of the logarithm of the half-lives of the Centaurs
with their semi-major axis, perihelion and aphelion distances, and
their inclination. Points in red show the results of the forward
direction, and those in green show the results of the backward
direction integrations.}
\label{fig:orbitlnhalf-life}
\end{figure*}

The half-lives of individual Centaurs are given in
Table~\ref{tab:half-lives}. The value of the Poisson uncertainty
$\sigma$ is calculated as $\sigma = {T_{1/2} / \sqrt{N_0}}$, where
$N_0$ is the initial number of clones. This uncertainty is added in
quadrature to the uncertainty in the fitted half-life, as judged from
the $\chi^2$ surface in the space of model parameters and as returned
by our fitting software.  Note that our ignorance of individual error
bars on our simulation datapoints may lead to an underestimate of the
latter quantity -- it is generally smaller than $\sigma$ by an order
of magnitude -- as our algorithm is tantamount to assuming a good fit
to the exponetial decay law. The object with the shortest half-life is
1996 AR20, a JN object, which has a half-life of approximately 540 kyr
in the forward integration and 590 kyr in the backward
integration. The object with the longest half-life is 2000 FZ53, a UE
object, with half-lives of 26.8 Myr (forwards) and 32.3 Myr
(backwards).

On comparison with the orbital elements (Table \ref{tab:orbits}), a
correlation can be seen between the position of a Centaur's orbit
within the Solar system and its half-life. The farther from the
Edgeworth-Kuiper Belt, the shorter the half-life. This is not
unexpected -- Jupiter is significantly more massive than Saturn, and
Saturn in turn is more massive than either Uranus or Neptune. So, the
farther from the Edgeworth-Kuiper Belt, the more massive are the
planets with which the Centaur interacts and the more frequently do
such encounters occur.  This effect is evident when the orbital
elements of the objects are plotted against the logarithm of the
half-life, as in Figure \ref{fig:orbitlnhalf-life}. There is only a
rough correlation with semi-major axis, but the data indicate a lower
bound to the half-life as a function of perihelion $q$ and an upper
bound as a function of aphelion $Q$. Specifically, we find that
\begin{equation}
0.392\,\exp ( 0.135\, q) \lta T_{1/2} \lta 0.064\, \exp (0.275\, Q).
\end{equation}
where $q$ and $Q$ are in au and $T_{1/2}$ is in Myr. This holds for
all the Centaurs in our sample, but it is conceivably possible that
low eccentricity orbits between the planets are extremely long-lived
(e.g., Holman 1997; Evans \& Tabachnik 1999). Additionally, in the
plot of perihelion distance versus half-life, there seem to be three
rough groupings of objects. The first are those in a band along the
line from $q = 6$, $\log T_{1 \over 2} = 5.75$ to $q = 26$, $\log T_{1
\over 2} = 7.2$, which accounts for the bulk of the objects. A second
group comprises six objects, which lie roughly on a parallel track at
values of $\log T_{1 \over 2}$ greater by 0.7.  Finally, 2000 FZ53
sits alone far above either of these groups. The objects which are in
the first group currently lie away from the positions of any major
resonances and so tend to have short lifetimes. The objects in the
second group tend to lie nearer to stable mean motion or secular
resonances. In fact, the orbit of 2002 GB10 lies within 0.009 au of
the 3:4 resonance of Uranus, while the orbit of Chariklo lies within
0.09 au of the 4:3 resonance of Uranus. Clones of 2000 FZ53 quite
frequently display resonant behaviour during the course of the
simulations, although it does not currently lie near any major mean
motion resonances. A possible cause of 2000 FZ53's exceptionally long
half-life is its abnormally large inclination -- higher than that of
any other Centaur studied by over 10 degrees. Any correlations of
half-life with inclination and ecccentricity are less clear-cut than
those with position. However, there is a lack of long-lived objects at
large $e$.  This is a consequence of the fact that Centaurs with large
$e$ must cross the orbits of several of the outer planets, and so
inevitably are more unstable than bodies whose close approaches are
restricted to just one or two planets.

As the equations of motion are time-reversible, it might naively be
expected that the forward and backwards half-lives should be the
same. In fact, it is often found that inward evolution of the orbits
of minor bodies is more likely than outward evolution (e.g., Oikawa \&
Everhart 1979).  This is because the Edgeworth-Kuiper belt provides a
source while Jupiter provides a sink, so that clones ejected by
Jupiter do not have an opportunity to return. Of all the objects, 19
have shorter half-lives in the forward direction, and 13 have shorter
half-lives in the backward direction. The least discrepancy occurs for
2001 BL41, with the two half-lives agreeing to within 0.25\%. The
greatest discrepancy occurs for 2002 GZ32, where the backward
half-life is 46\% longer than the forward half-life. These differences
can be visualised as the effect of the first encounters the clones
have with the major planets. If an encounter with one of the planets
occurs early enough in the simulation, the entire ensemble of clones
will be perturbed, moving the objects onto slightly different orbits,
with slightly different half-lives.  For the entire dataset, the
forward and backward half-lives only diverge by a matter of 30000
years out of 2.76 Myr -- a discrepancy of only just over 1\%.

\begin{table}
\begin{center}
\begin{tabular}{||c|c|c|c|c||} 
 \hline
 \hline
 C   & Direction & $N$ & $T_{1 \over 2}$ & $\sigma$ \\ \hline \hline
 JS  & Forward & 729 & 0.67 & 0.03 \\ 
 \null & Backward & 729 & 0.65 & 0.02 \\ \hline

 JU  & Forward & 729 & 0.61 & 0.02 \\
 \null & Backward & 729 & 0.64 & 0.02 \\ \hline

 JN  & Forward & 891 & 0.59 & 0.02 \\ 
 \null & Backward & 891 & 0.63 & 0.02 \\ \hline

 S  & Forward & 729 & 0.72 & 0.03 \\
 \null & Backward & 729 & 0.68  & 0.03  \\ \hline

 SU & Forward & 3519 & 1.03 & 0.02 \\
 \null & Backward & 3519 & 1.02 & 0.02 \\ \hline

 SN & Forward & 2205 & 1.11 & 0.02 \\
 \null & Backward & 2205 & 1.22 & 0.03 \\ \hline

 SE & Forward & 1917 & 1.75 & 0.04     \\
 \null & Backward & 1917 & 1.77 & 0.04 \\ \hline

 U  & Forward & 3780 & 3.72 & 0.06 \\
 \null & Backward & 3780 & 3.36 & 0.06 \\ \hline

 UN & Forward & 5094 & 5.84 & 0.08 \\
 \null & Backward & 5094 & 6.19 & 0.09 \\ \hline

 UE & Forward & 3006 & 12.5  & 0.23 \\
 \null & Backward & 3006 & 14.2  & 0.26 \\ \hline

 N  & Forward & 729  & 12.5  & 0.46 \\
 \null & Backward & 729 & 13.5 & 0.50 \\ \hline \hline

 \end{tabular}
\end{center}
\caption[Half-Lives of the classification bins]
{\label{tab:binhalf-lives} The half-lives of the individual
classification bins across which the objects fall {\it at the start}
of the integrations. $N$ is the number of clones in that particular
class at the start of the integrations, $T_{1 \over 2}$ is the
half-life (in Myr), and $\sigma$ is again the uncertainty on the
half-life (in Myr).}
\end{table}

From the dataset, it is also possible to calculate the half-lives of
the starting class of the objects.  The results of this calculation
are given in Table \ref{tab:binhalf-lives}. The number of clones in
any particular class is not necessarily an exact multiple of 729. This
is because a number of the objects have outlying clones which actually
fall into a different class at the start of the integration as
compared to the seed. The results of this analysis again show the
dependence of half-life on perihelion position -- objects in the
Jupiter classes have half-lives shorter than those in the Saturn
classes, and these in turn are more short-lived than the objects in
the Uranus classes. There are also hints in the table that more
eccentric objects under the control of any particular planet may be
more long-lived than their less eccentric counterparts (compare, for
example, the half-lives of U, UN and UE objects). Orbital periods of
the more eccentric objects are greater than those on near-circular
orbits with similar perihelion distances, and hence encounters with
the giant planets happen less frequently.

Over the course of the integrations, the clones of each object are
repeatedly transferred between classification bins.  This allows us to
evaluate the amount of time that is spent in each of these classes
over the simulation, together with the number of times the object is
transferred into that class. From this, we can calculate the mean time
that an object spends in any particular class before being transferred
into another.  The results of such calculations are presented in Table
\ref{tab:binlifetimes}. In this table, the value of the mean lifetime
for the L or long-period comet class has been ignored, since objects
which enter this classification are then removed from the simulation
as they pass 1000 au from the Sun. This means that the value of mean
lifetime for objects of class L is unrealistically small. It is also
noteworthy that the EK and T classes have particularly short mean
lifetimes. A stable orbit in these regions requires decoupling from
Neptune, and there are no non-gravitational forces within the
integrations which could allow this to happen. Hence, the very small
number of objects which attain these two classes only do so at the
extremes of a series of perturbations and are immediately perturbed
back into classes under the control of Neptune.

\begin{table}
\begin{center}
\begin{tabular}{||c|c|c||} 
 \hline
 \hline
Class & Forward lifetime & Backward lifetime \\ \hline \hline
EC    &	1960 & 1990  \\ \hline
MC    &  950 &  890  \\ \hline \hline
E     & 4020 & 3480  \\ \hline
SP    & 1680 & 1630  \\ \hline
I     &  570 &  630  \\ \hline
J     &  290 &  300  \\ \hline
JS    &  890 &  890  \\ \hline
JU    & 1320 & 1310  \\ \hline
JN    & 1110 & 1120  \\ \hline
JE    & 1470 & 1470  \\ \hline
JT    &  830 &  860  \\ \hline
S     & 1400 & 1350  \\ \hline
SU    & 3670 & 3630  \\ \hline
SN    & 2710 & 2710  \\ \hline
SE    & 3060 & 3010  \\ \hline
ST    & 2960 & 3090  \\ \hline
U     & 6650 & 6150  \\ \hline
UN    & 4860 & 5040  \\ \hline
UE    & 4710 & 4870  \\ \hline
UT    & 1560 & 1480  \\ \hline
N     & 2600 & 2500  \\ \hline
NE    & 4640 & 4350  \\ \hline
NT    & 4100 & 4440  \\ \hline
EK    &  330 &  320  \\ \hline
T    &  100 &  100  \\ \hline \hline
\end{tabular}
\end{center}
\caption[Mean time spent in each classification bin]
{\label{tab:binlifetimes} The mean time (in yrs) spent in each
classification bin, before a Centaur clone is transferred to another
bin. The shortness of this mean time is understandable, as objects
close to the boundary are often tranferred to and fro. Note that EC
stands for Earth-crossing objects and MC for Mars-crossing.}
\end{table}

\section{Transfer Probabilities}

It is also straightforward to calculate the probability of an object
being transferred from one class to another. This can be visualised by
constructing a $24 \times 24$ grid with the initial class on the
vertical axis and the final class on the horizontal (these classes are
just initial and final with respect to a single transfer, not for the
entire integration). This is done by recording every transfer which
occurs within the integrations, and hence calculating the fraction of
objects which, for example, are transferred from class J to class
JS. The results are shown in Table~\ref{fig:BackwardTransfers}. The
numbers have been normalised so that the sum along any row is
unity. For any class, the probabilities give the relative likelihood
of leaving from that class to the target classes given on the
horizontal axis. As an example, let us take a typical result from one
of these tables, namely that the value of the probability of transfer
from class J to class JS is $\sim 0.49$ (Table
\ref{fig:BackwardTransfers}). This means that, for an object in the J
class, there is a $\sim 49\%$ chance of the object being transferred
directly into the JS class the next time the classification
changes. We can see that for such an object, the two most likely
transfers are to the JS class or to a short-period (SP) comet, and
between them, these two possibilities make up the great bulk of
transfers for all J objects.

Table~\ref{fig:BackwardTransfers} shows a number of interesting
features. Whenever an object is controlled by two planets (one at
perihelion and one at aphelion), the classes to which the object is
most likely to move correspond to transitions at the perihelion and
aphelion of the planet. For instance, an SU object is controlled at
perihelion by Saturn and at aphelion by Uranus. It is most likely to
be transferred to one of the classes JU or U by an encounter at
aphelion.  These cases corresponds to an encounter with Uranus either
increasing the eccentricity of the orbit, and hence pushing the
perihelion down to Jupiter's control, or decreasing the eccentricity,
pulling the perihelion away from Saturn's control.  For encounters at
perihelion, the most likely classes are S or SN corresponding either
to a circularisation of the orbit at Saturn, or to a pumping of the
eccentricity of the orbit, as the aphelion moves from Uranus' to
Neptune's control.  These most popular transfers can be traced
diagonally down the tables, around the empty diagonal corresponding to
the same initial as final class. These four parallel lines of high
probabilities give the appearance of two sets of ``tram lines''
running through the tables.  After these possibilities, other
transfers are also viable, albeit with lower probabilities - for
example, an SU object can suffer a perihelion-aphelion interchange at
Saturn, moving to the JS class. However, the fact that the four
classes most likely to be reached in a transfer lie along the ``tram
lines'' vindicates the classification scheme, which is based on the
idea of transfers by interaction primarily at perihelion and aphelion.

Also of interest is an effect which can be seen on comparing Table
\ref{fig:BackwardTransfers} with the equivalent results for each
individual Centaur (given in Appendix A of Horner 2003).  From any
class, the probability of transfer to another class is roughly
constant, regardless of the direction of integration or the object in
question. The main discrepancies lie in very low probability
transfers, where the uncertainty is large because of the small numbers
involved. This means that for a newly discovered object, it is
possible to give the probabilities of its transfer to any new class,
as long as the initial class can be computed. It also permits insight
into the main dynamical pathways followed by a Centaur. For example,
using the values in Table ~\ref{fig:BackwardTransfers} and assuming an
initial population of 1000 short-period comets or SP objects, it can
be seen that 27 of these objects become E types, all of which would
return to the SP bin, 41 enter the I class, three of which would on
average then return, 304 travel to the J class, 147 making the return
trip. For the other bins, we find 84 returning objects from JS, 8 from
the JU class and one from JN. Therefore, 270 of 1000 objects that
leave the short-period class return immediately the next time that
their classification changes.

Given that the total time spent in each class and the number of times
that class becomes occupied are calculable from the simulation data,
it is also possible to compute the probability per unit time of a
transfer.  We already know the probabilities that an object in a
particular class will be transferred to any other. Dividing the mean
time spent in any class by this probability, we obtain the mean
transfer time from one class to any target class. The inverse of this
is the probability per unit time of the transfer.  The values obtained
in this type of analysis are given in Horner (2003) for the individual
Centaurs, while the probabilities for the entire dataset are given in
Table~\ref{fig:Backwardtimes}. The results are given as probabilities
per Myrs, so that a value of 0.1 in a particular box means that an
object making the relevant transfer would have a mean transfer time of
about 10 Myrs. This means that the population in any class $N$ evolves
according to
\begin{equation}
{d N \over dt} = \sum_i P_i N_i - \sum_j P_j N_j
\end{equation}
where $P_j$ and $N_j$ are the probabilities per unit time and the
populations in the bins along the row, while the $P_i$ and $N_i$ are
the corresponding quantities along the column. In other words, the
ingress to a particular class is governed by the numbers in the
column, and the egress by the row. Mathematically speaking, this gives
us coupled sets of exactly solvable linear first-order differential
equations that govern the evolution of Centaur clones. We will return
to this in a later publication, but these ideas are already prefigured
in Bailey et al. (1992).

\begin{table*}
\psfig{figure=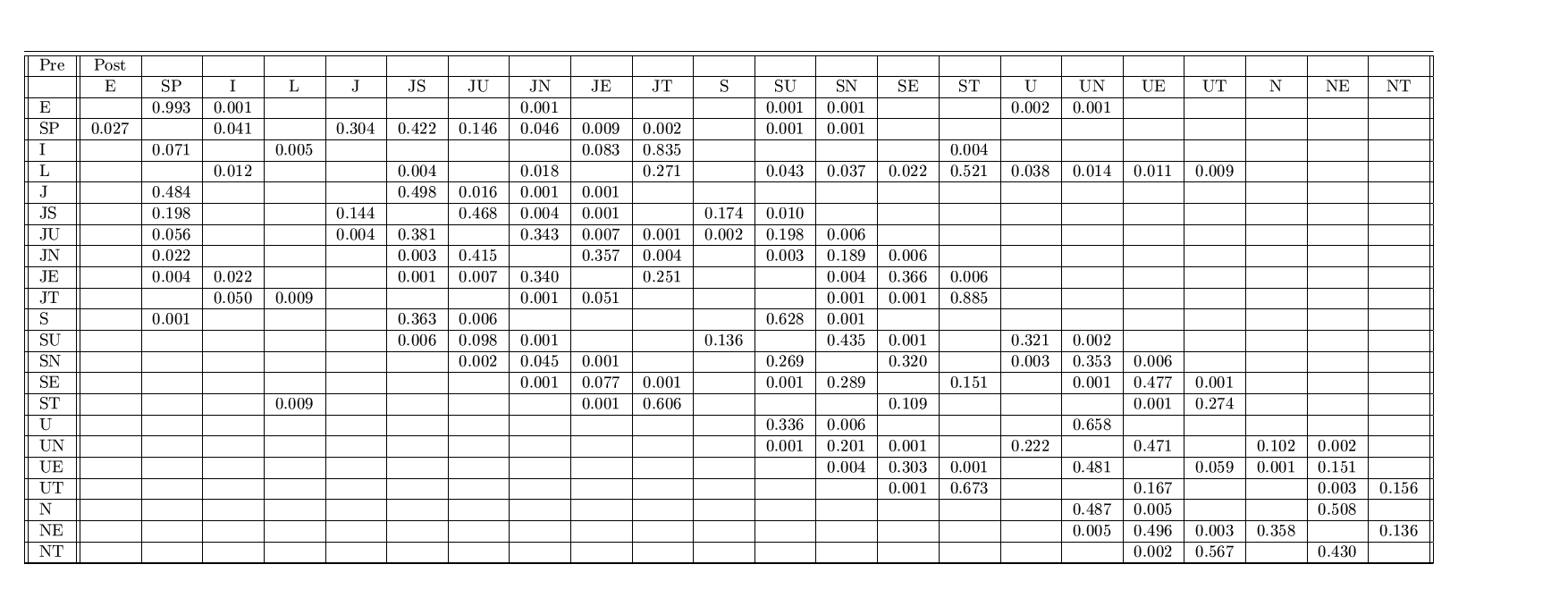,height=10cm,width=19.5cm,angle=90}
\psfig{figure=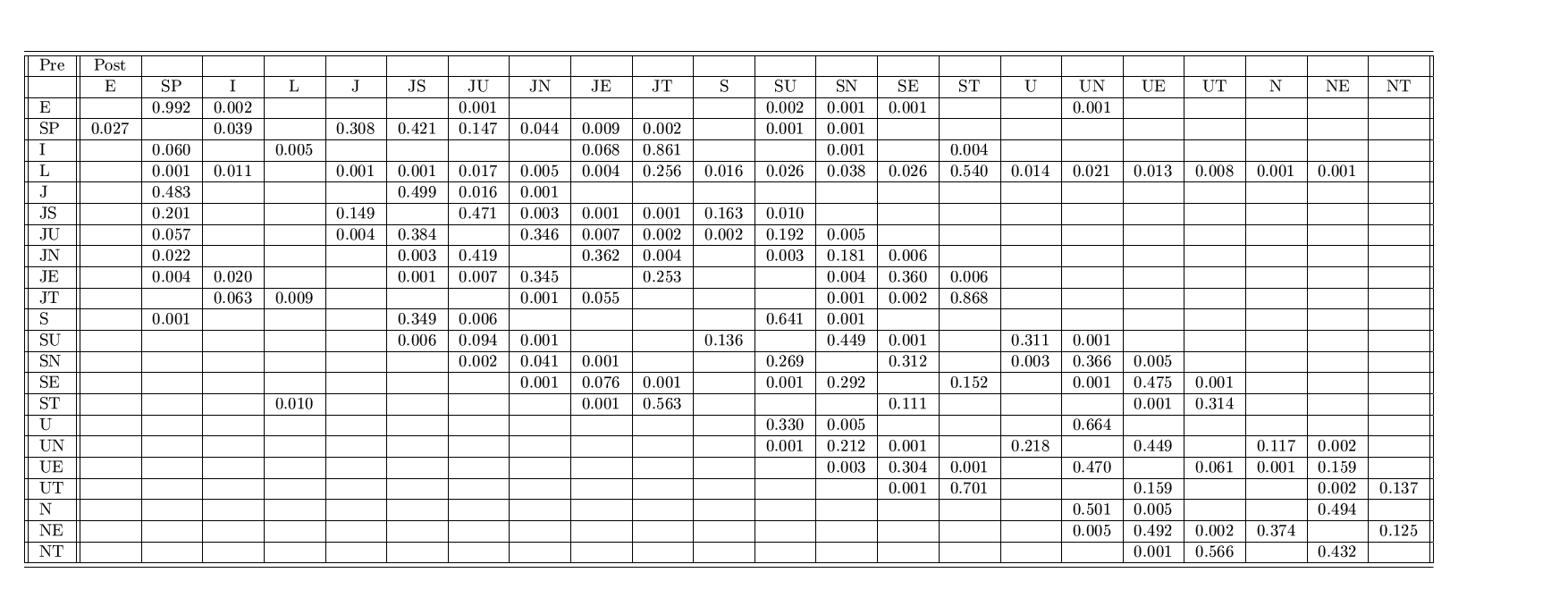,height=10cm,width=19.5cm,angle=90}
\caption{Transfer probabilities for the entire simulation in the
forward (upper table) and backward (lower table) direction. Notice
that the probabilities are nearly the same regardless of the direction
of integration. An empty entry means that the transfer probability is
$< 10^{-3}$. The leading diagonal is empty by definition.}
\label{fig:BackwardTransfers}
\end{table*}
\begin{table*}
\psfig{figure=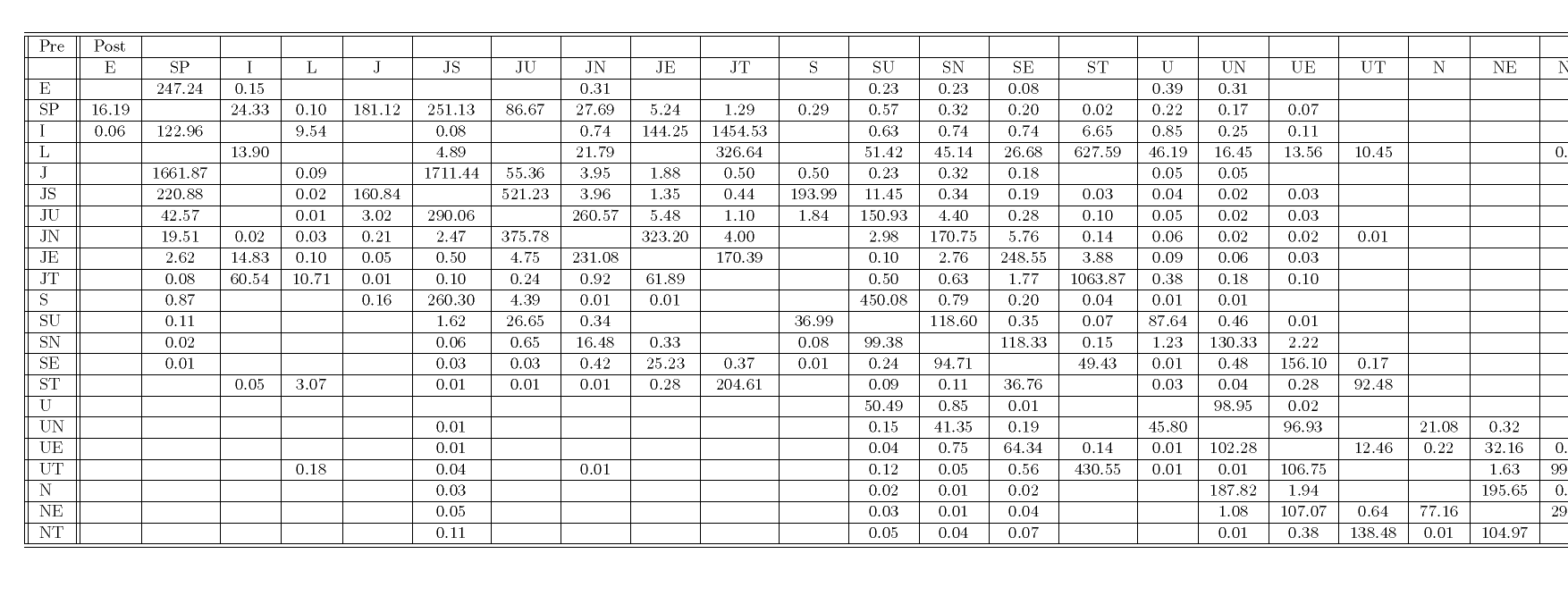,height=10cm,width=17.5cm,angle=90}
\psfig{figure=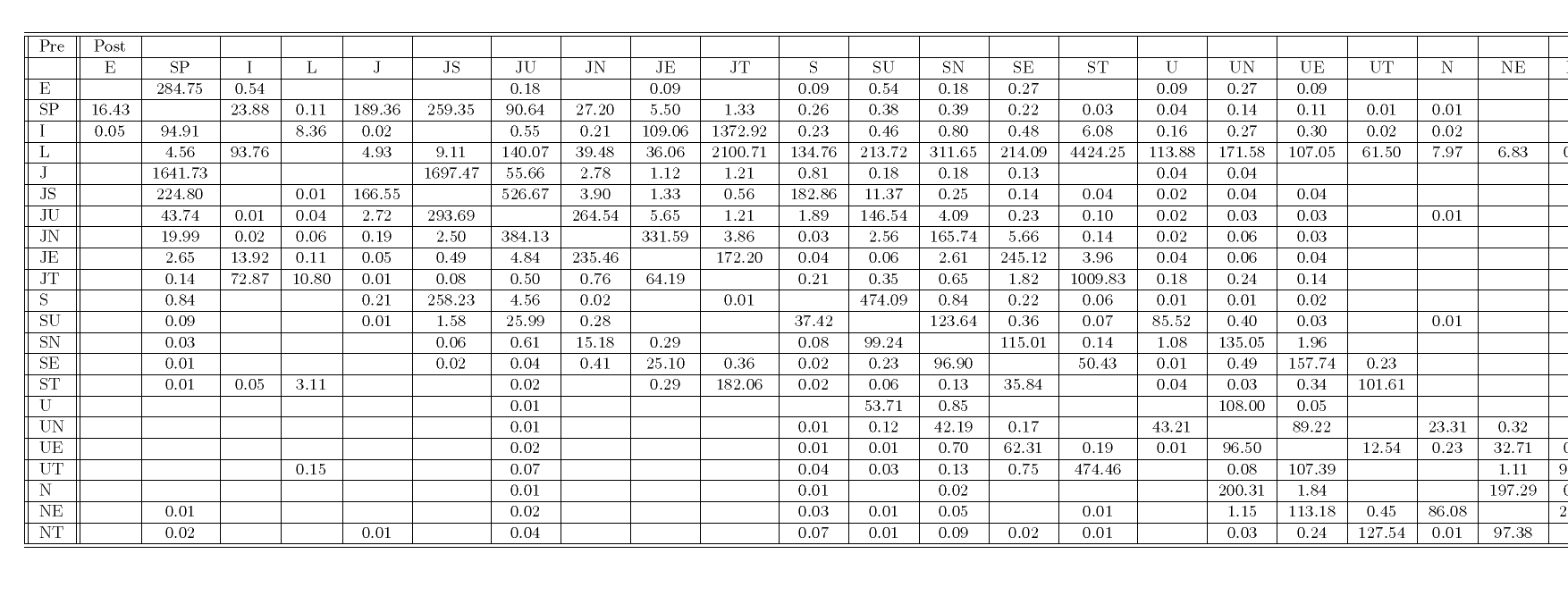,height=10cm,width=17.5cm,angle=90}
\caption{Transfer probabilities per Myr for the entire simulation in
the forward (upper table) and backward (lower table) direction.  An
empty entry means that the transfer probability per Myr is $<
10^{-2}$. The leading diagonal is empty by definition.}
\label{fig:Backwardtimes}
\end{table*}

\section{Centaur Fluxes and Population}

During the simulations, we also record the numbers of clones that
become Earth-crossing objects, Mars-crossing objects and short-period
comets. This gives us a means to reckon the total population of
Centaurs.  Fernandez (1985) suggested a flux of $10^{-2}$ new
short-period comets per year, with a mean lifetime of $\sim 6$
kyr. More recently, Levison \& Duncan (1994) find that the mean fade
time for a short-period comet is $\sim 12$ kyr. Under the assumption
that the current population of short-period comets is in steady state,
the work of Levison \& Duncan implies a flux of $\sim 0.5 \times
10^{-2}$ new short-period comets per year. This is equivalent to one
new short-period comet being captured, on average, every 200 years. If
we assume that the entirety of this flux comes from the Centaur
region, then this allows us to estimate the total population of the
Centaur region using the simulations.

From the simulation data, a total of 7\,900 out of 23\,328 clones
($\sim 34\%$ of the initial population) become short-period comets at
some point during the forward integrations, and 8\,068 (again $\sim
34\%$) become short-period comets during the backward integrations.
This is a flux of one new short-period comet every 380 years. If we
assume all short-period comets are captured from the Centaur region,
an estimate of the total number of Centaurs (with perihelion distance
$q \gta 4$ au and aphelion distance $Q \lta 60$ au) is $\sim$ 44\,300.
This represents the population of objects bright enough to be visible
as short-period comets, were they to be captured into such an
orbit. An effective nuclear diameter $d$ greater than 1 km seems a
reasonable limit to place for this value, though there are an
increasing number of comets with $d \sim 0.5$ km (e.g., Lamy et
al. 2004). This calculation also only takes into account objects
becoming short-period comets for the first time. Given that the mean
fade time is $\sim 12$ kyr, it is reasonable to assume that objects
captured for the first time are significantly brighter as short-period
comets than those which have experienced a number of prolonged stays
in the region. Therefore, the objects, which in the simulations
display a number of prolonged periods as short-period comets, would
actually exhaust all their volatiles early on and should not
contribute to the new short-period comet flux in later passages
through the inner Solar system. Note that the usage of the flux of
short-period comets to normalise the source populations has also been
exploited recently by Emel'yanenko, Asher \& Bailey (2004) to estimate
the total population of trans-Neptunian objects on highly eccentric
orbits, which reside at still greater heliocentric distances than the
Centaurs.

Given our estimate of the total population of Centaurs and the
knowledge that the half-life is $\sim 2.75$ Myr, we can estimate the
influx of new Centaurs from the Edgeworth-Kuiper belt. Neglecting
those few objects that could be captured onto Centaur-like orbits from
high eccentricity orbits from the Oort cloud, we can see that $\sim
22150$ objects must be replaced every 2.75 Myr. This is equivalent to
1 object transferred to the Centaur region from the Edgeworth-Kuiper
belt every $\sim 125$ yrs. This calculation also ignores the small
flux of objects from Centaur-type orbits to the Edgeworth-Kuiper
belt. Our simulations allow neither for non-gravitational effects,
such as collisions, nor for the gravitational perturbations between
Edgeworth-Kuiper belt objects. So, it is impossible to determine the
flow of objects from orbits which encounter Neptune to those which are
stable beyond.  To drive the inward flux of Centaurs, the effects of
collisions and of perturbations between the Edgeworth-Kuiper belt
objects must be considered. Durda \& Stern (2000) suggest that
collisions of objects greater than 4 m in diameter onto comet-sized
bodies within the Edgeworth-Kuiper belt occur every few days. In
addition, they reckon that the time scale for the disruption of 1 km
objects is $\sim 1$ Gyr. These two facts in concert imply a high rate
of collision within the Edgeworth-Kuiper belt, sufficiently high that
an inward flux of 1 new Centaur every $\sim 125$ yrs seems reasonable.

From the forward integrations, we can deduce that 1799 clones (7.7\%
of the sample) became Earth-crossing and 3799 clones (16\%) became
Mars-crossing. Very similar numbers are yielded by the backward
integrations. Therefore, we expect typically one Centaur to become
Earth-crossing for the first time approximately every $\sim 880$ yrs,
and one new Mars-crosser every $\sim 420$ yrs.  Most of the known
population of Near-Earth objects (NEOs) is asteroidal in nature, as
the Main Belt provides the great majority of NEOs. Morbidelli et
al. (2002) state that only 6\% of the NEOs are ultimately of
Edgeworth-Kuiper Belt origin. NEOs originating in the Main Belt are
expected to survive for far longer times within the inner Solar system
than the cometary bodies, due to the fact that they lie on orbits
significantly decoupled from direct perturbations by Jupiter. An
approximate flux of one new Earth-crosser of Centaur (and hence,
originally Edgeworth-Kuiper belt) origin per millennium seems to be in
reasonable agreement with such work.

Though the number of NEOs of Edgeworth-Kuiper belt origin is small,
they are exceptionally important in judging potential hazards.  The
size distribution of NEOs of asteroidal origin is heavily weighted
towards small particles, consistent with the idea that they are
collision fragments. The largest NEO is (1036) Ganymed with an
absolute magnitude $H = 9.45$ and a diameter $d \approx 50$ km.  Very
few NEOs are larger than $10$ km across.  However, for the Centaurs,
the upper end of the size distribution is well-populated, with 16
objects of the 32 listed in Table~\ref{tab:orbits} intrinsically
brighter than Ganymed. The passage of a large Centaur like Chiron or
Pholus into the inner Solar system would provide a very significant
environmental disturbance (Hahn \& Bailey 1990), as its fragmentation
and possible decay could overwhelm the local space environment with
debris and dust.

Another interesting set of statistics is the total number of times a
class becomes occupied during the integrations. In the forward
direction, the Earth-crossers become occupied $\sim$ 23\,300 times over
three million years. However, we already know that only 1799 of the
clones enter this class, so it is obvious that the bulk of objects
which become Earth crossing at least once will in fact enter, leave
and re-enter the class a number of times (in this case, $\sim 13$
times). Table \ref{tab:Visits} summarises the number of visits to the
short-period (SP), Earth-crossing (EC), Mars-crossing (MC) and
Encke-type (E) classes that a typical object entering these classes
would have.

 \begin{table}
 \begin{center}
 \begin{tabular}{||c|c|c||}
 \hline \hline
 Class & D & N \\ \hline \hline
 EC    & F & 12.9 \\
 \null & B & 13.7 \\ \hline
 MC    & F & 13.7 \\ 
 \null & B & 14.6 \\ \hline
 SP    & F & 28.0 \\ 
 \null & B & 28.4 \\ \hline
 E     & F & 16.5 \\ 
 \null & B & 18.3 \\ \hline \hline

 \end{tabular}
 \caption[Mean number of visits to cometary classes]
{ \label{tab:Visits}
 Table showing the mean number of times an object which enters a
 cometary class at least once will go on to enter and re-enter that
 class through its lifetime. D gives the direction of the integration
 (F = Forward, B = Backward) while N gives the mean number of visits
 to that class. Here, again EC is Earth-crossing, MC is Mars-crossing,
 SP is short-period and E is Encke-type.} 
 \end{center}
 \end{table}

The flux of objects into the Encke class of comets can also be
calculated. This is subject to greater errors -- because of the much
lower numbers of objects that become Encke-type comets than become
Earth-crossers and because a time step of 120 days is insufficient to
track the orbits of such objects with reasonable accuracy and because
the gravitational influence of the neglected terrestrial planets is
now significant.  For objects in the forward integrations, 303 clones
became Encke types, implying a capture rate of one every $\sim 5200$
yrs. The backward integration yields the same capture rate.

Typically, an object will become an Earth-crosser or Mars-crosser
while classified as a short-period comet, having been transferred to
the inner Solar system by an encounter with Jupiter. Although some
objects can be perturbed onto Earth and Mars crossing orbits by the
other outer planets (for example, the comets P/1997 T3
Lagerkvist-Carsenty and P/1998 U3 J\"ager, discussed in Horner et
al. 2003), the vast majority of such objects are under the control of
Jupiter prior to a period of cometary behaviour.  This is evident in
Table~\ref{fig:BackwardTransfers}. In both the forward and backward
integrations, the only classes from which the likelihood of transfer
to short-period orbits is greater than 0.001 are the other cometary
classes (E, I, L), the classes in which the object is controlled by
Jupiter (J, JS, etc.), and the S class.

Finally, we can use the simulation data to estimate the impact rate on
the giant planets from Centaurs.  In the forward integration, we find
that 144 objects hit Jupiter, 53 hit Saturn, 5 hit Uranus and a
further 5 hit Neptune. In the backward integration, these numbers are
135, 48, 5 and 1 respectively.  Given that the estimated population of
the Centaur region is $\sim 44300$, then we expect one impact per 10
kyr on Jupiter, one per 28 kyr on Saturn, and one per 300 kyr on
Uranus and Neptune. These numbers are likely to be underestimates of
the impact rate on the planets, given that the errors in integration
are at their largest when the clone is closest to a massive body.

Of course, all the numbers derived in this Section are dependent on
the assumed flux of one new short-period comet every 200 years.  If
the true flux is higher or lower, then the total populations of
Centaurs, Earth-crossers and Mars-crossers would need to be
correspondingly adjusted upwards or downwards. Although the flux of
new short-period comets is perhaps uncertain by a factor of two, it is
not uncertain by a factor of ten, and so our population estimates are
surely correct to an order of magnitude.

 \begin{table*}
 \begin{tabular}{||c|c|c|c||c|c|c||}
 \hline
 Object & Class & D & $T_{1 \over 2}$ & \null & Colour & \null \\ 
 \null & \null & \null & \null & B - V & V - R & R - I \\ \hline

 2000 EC98      & JU & F &  0.61  & \null & $0.47\pm 0.06$ & $0.50 \pm 0.06$ \\ 
 \null      & \null & B  &  0.63 & $0.854 \pm  0.081$ &  $0.466 \pm 0.05$ & $ 0.439 \pm 0.076 $ \\ \hline

 1998 SG35      & JS & F   &  0.67  & & $0.42 \pm 0.08$ & $0.46 \pm 0.04$ \\ 
 \null      & \null & B  &  0.65 & $0.725 \pm  0.089$ & $0.456 \pm 0.050$ & $0.546 \pm 0.063$ \\ \hline

 1999 UG5       & SU & F   &  0.74  & $0.88 \pm 0.18$ & $0.60 \pm 0.08$ & $0.72 \pm 0.13$ \\ 
 \null       & \null & B  &  0.85  & $0.964 \pm 0.085$ & $0.607 \pm 0.060$ & $0.625 \pm 0.042$ \\ 
 \null          & \null & \null & \null & \null & $0.68 \pm 0.02$ & $0.58 \pm 0.02$ \\ \hline      

 Asbolus   	& SN & F   &  0.86   & $0.75 \pm 0.04$ & $0.47 \pm 0.04$ & \\ 
 \null  	& \null & B  &  0.75  & $0.750 \pm 0.040$ & $0.513 \pm 0.068$ & $0.523 \pm 0.045 $\\ 
 \null          & \null & \null & \null & \null & $0.53 \pm 0.02$ & $0.47 \pm 0.02$ \\ \hline      

 2001 PT13        & SU & F   &  0.94  & & & \\ 
 \null            & \null & B  &  0.87 & \null & $0.44 \pm 0.03$ & $0.48 \pm 0.02 $ \\ \hline 

 2001 BL41        & SU & F   &  0.95  & & & \\ 
 \null            & \null & B  &  0.95 & \null & $0.53 \pm 0.05$ & $ 0.56 \pm 0.04 $ \\ \hline 

 Chiron    	& SU & F   &  1.03  & $0.67 \pm 0.06$ & $0.34 \pm 0.03$ &\\ 
 \null    	& \null & B  &  1.07  & $0.679 \pm 0.039$ &$0.359 \pm 0.027$ & $0.356 \pm 0.037$ \\ 
 \null          & \null & \null & \null & \null & $0.36 \pm 0.10$ & $0.32 \pm 0.10$ \\ \hline      

 1999 XX143        & SN & F   &  1.06  & & & \\ 
 \null        & \null & B  &  1.38 & \null & $0.67 \pm 0.07$ & $ 0.70 \pm 0.06 $ \\ \hline 

 Pholus    	& SN & F   &  1.28  &  $1.35$ & $0.71$ & \\ 
 \null    	& \null & B  &  1.39  & $1.19 \pm 0.10$ & $0.78 \pm 0.04$ & $0.81$ \\ 
 \null     	& \null & \null      &  \null  & $1.299 \pm 0.099$ & $0.794 \pm 0.032$ & $0.814 \pm 0.056$ \\ \hline

 1994 TA        & SU & F   &  1.78  & & & \\ 
 \null        & \null & B  &  1.52 & $1.261 \pm 0.139$ & $0.672 \pm 0.080$ & $ 0.740 \pm 0.210 $ \\ \hline 

 2002 GO9        & UN & F   &  2.93  & & & \\ 
 \null        & \null & B  &  3.67 & \null & $0.74 \pm 0.06$ & $ 0.66 \pm 0.05 $ \\ \hline 

 2000 QC243     & U & F   &  3.18  & & $0.38 \pm 0.06$ & $0.41 \pm 0.06$ \\ 
 \null     & \null & B  &  3.44 & $0.724 \pm 0.062$ & $0.448 \pm 0.044$ & $0.397 \pm 0.069$ \\ \hline 

 1998 QM107     & UN & F   &  4.87  & $0.771 \pm 0.100$ & $0.474 \pm 0.095$ & $0.368 \pm 0.102 $ \\ 
 \null     & \null & B  &  5.65  & $0.730 \pm 0.060$ & $0.520 \pm 0.030$ & - \\ 
 \null          & \null & \null & \null & \null & $0.63 \pm 0.12$ & $0.64 \pm 0.10$ \\ \hline      

 Nessus    	& SE & F   &  4.91  & $0.88 \pm 0.07$ & $0.77 \pm 0.05$ & \\ 
 \null    	& \null & B  &  6.40  & $1.090 \pm 0.040 $ & $0.793 \pm 0.041$ & $0.695 \pm 0.066$ \\ 
 \null          & \null & \null & \null & \null & $0.74 \pm 0.08$ & $0.64 \pm 0.07$ \\ \hline      

 Hylonome  	& UN & F   &  6.37  & & $ 0.41 \pm 0.10 $ & $ 0.46 \pm 0.18 $ \\ 
 \null  	& \null & B  &  7.30  & $0.643 \pm 0.082$ & $0.464 \pm 0.059 $ & $0.490 \pm 0.122$ \\
 \null          & \null & \null & \null & \null & $0.50 \pm 0.07$ & $0.52 \pm 0.06$ \\ \hline      

 2002 GB10        & UE & F   &  11.1  & & & \\ 
 \null        & \null & B  &  13.1 & \null & $0.58 \pm 0.07$ & $ 0.62 \pm 0.07 $ \\ \hline 

 Chariklo  	& U & F   &  10.3  & & $~ 0.47$ & $0.55$ \\ 
 \null  	& \null & B  &  9.38    & $0.802 \pm 0.049$ & $0.479 \pm 0.029$ & $0.542 \pm 0.030 $ \\ 
 \null          & \null & \null & \null & \null & $0.49 \pm 0.02$ & $0.51 \pm 0.02$ \\ \hline      

 1998 TF35      & UE & F   &  11.5       & & $0.65 \pm 0.08$ & $0.66 \pm 0.07$ \\ 
 \null          & \null & B  &  10.8  & $1.085 \pm 0.111 $ & $0.697
 \pm 0.064$ & $0.651 \pm 0.119$ \\ \hline

 2002 GB10 & UE & F & 11.1 & & & \\ \null & \null & B & 13.1 & \null &
 $0.71 \pm 0.02$ & $ 0.65 \pm 0.02 $ \\ \hline 
 \end{tabular}
 \caption[Centaur colours and halflives] {\label{tab:Colours} List of
 the Centaurs with colour information. D is the direction of
 integration and $T_{1 \over 2}$ is the half-life in Myr. The colours
 are taken from Hainaut \& Delsanti (2002), Boehnhardt et al. (2002)
 and Bauer et al. (2003).}

\end{table*}

\section{Dynamics and Photometric Colours}

Since the Centaurs range over a large area of the outer Solar system,
it is possible that there could be variations in their observable
characteristics (such as colour and light-curves) as a function of
their position. Observations have been carried out by a number of
groups (e.g. Hainaut \& Delsanti 2002; Peixinho et al. 2001; Weintraub
et al. 1997; Bauer et al. 2003). Although the number of Centaurs with
well-determined colours is restricted, the situation is rapidly
improving. Colours for 19 Centaurs are given in Table
\ref{tab:Colours}. They range from some of the reddest objects in the
Solar system (e.g., Pholus) to much bluer objects (e.g.,
Chiron). There appears to be no correlation between colour and
half-life within the data set, or indeed between colour and
classification (for example the reddest and the bluest objects on the
list, Pholus and Chiron, are both controlled by the same planet at
perihelion).

We expect objects closest to the Sun to have undergone some
outgassing, and this would lead to a re-surfacing of the object,
covering the older, darker and redder material with fresh material
from the interior of the object. Conversely, those objects which have
not displayed cometary behaviour since the formation of the Solar
system are expected to be darker and redder.  Other factors that could
alter the colours of the Centaurs include impacts between objects.
Collisions might activate areas of the surface of the objects,
exposing fresh material from the interior, manifesting itself as bluer
colours. It follows that photometric observations of the Centaurs do
have the potential to provide important clues as to the dynamical
history of the object which cannot be determined by integration alone.

A good example of this is the case of Chiron and Pholus. These two
objects have similar half-lives, and are controlled by the same
planet. We can easily calculate that $\sim 60 \%$ of the clones of
Chiron become short-period comets in the forward integrations, against
$\sim 30 \%$ of the clones of Pholus. Yet, it is impossible to
determine whether either object has been a short-period comet in the
past, purely by means of the integrations. All that we can calculate
are percentage probabilities. However, it is possible that the fact
that Chiron is the bluest of the Centaurs listed in Table
\ref{tab:Colours} is caused by its well-known cometary activity, which
may indicate that at some time in the past Chiron was a short-period
comet. Pholus' red colour may similarly indicate it has not yet been
transferred into an orbit which would activate the surface
sufficiently for resurfacing to take place. Hence, we may be observing
the surface of an object which has not been active since the birth of
the Solar system.

Bauer et al. (2003) find that the the Centaurs display a wider colour
distribution than both Jupiter-family short-period comets and
Edgeworth-Kuiper belt objects. The members of the former population
are all active, while the members of the latter are not. An
intermediate population such as the Centaurs should at least show
properties varying between one extreme and the other, and this seems
borne out by the data in Table~\ref{tab:Colours}.

\section{Conclusions}

Detailed simulations of the evolution of the Centaur population under
the gravitational influence of the Sun and the 4 giant planets
(Jupiter, Saturn, Uranus and Neptune) have been carried out. 23\,328
test particles were created by taking the orbits of 32 well-known
Centaurs and producing 9 clones in each of semi-major axis ($a$),
eccentricity ($e$) and inclination ($i$), giving a total of 729 clones
of each object. The clones were then integrated in both the forward
and backward directions for a period of 3 Myr. Any clone which reached
a heliocentric distance of 1000 au was deemed ejected from the Solar
system and removed from the integration.

The half-lives for ejection of the Centaurs were calculated from the
simulation data. These ranged from $\sim 540$ kyr for 1996 AR20 (in
the forward direction) to $\sim 32$ Myr 2000 FZ53 (in the backward
direction). The half-life of the ensemble of Centaurs was $\sim 2.7$
Myr, irrespective of the direction of integration.  To analyze the
simulation data, we exploited a classification scheme introduced by
Horner et al. (2003), which breaks down the orbits of objects in the
Solar system into different classes according to the planet's
controlling the perihelion and aphelion. The use of the new
classification scheme allowed us to determine the dynamical pathways
through which the Centaur population evolves. Transfer probabilities
were calculated between the different classes. These were found to be
remarkably independent of the object under study, offering the
prospect of determining possible future histories for objects from
the knowledge of their current classification.

Both observations and simulations suggest that one new short-period
comet is produced every 200 yrs. This is used to normalise our
simulation results. In this case, the total population of the Centaur
region (from $q \gta 4$ au to $Q \lta 60$ au) with nuclei large enough
to provide visible comets ($ d \gta 1$ km) is estimated to be $\sim$
44\,300. A flux of 1 new object into the Centaur region from the
Edgeworth-Kuiper belt every 125 yrs is required to maintain the
population in a steady-state. Additionally, one fresh Earth-crossing
object is expected to arise from the Centaur region every $\sim 880$
yrs. This is both of interest and concern, as large Centaurs entering the
inner Solar system are likely to fragment with production of much
dangerous dust and debris.

In a companion paper (Horner et al. 2004), individual Centaur clones
are discussed, showing examples of stable resonant behaviour, Kozai
instabilities, capture into satellite orbits, and evolution into
objects spending long time periods in the inner Solar system, amongst
other things. This gives a feel for the rich variety of behaviour of
the diverse objects known as ``Centaurs''.

\section*{Acknowlegments}
JH acknowledges financial support from the Particle Physics and
Astronomy Research Council. We thank the referee S.C. Lowry for a
careful reading of the manuscript.


\label{lastpage}

\begin{thebibliography}{99}

\bibitem[Asher \& Steel(1993)]{1993MNRAS.263..179A} Asher D.J., Steel 
D.I.\ 1993, MNRAS, 263, 179 

\bibitem[Bailey(1993)]{} Bailey M.~E., Chambers J.~E., Hahn G.,
Scotti J., Tancredi G. 1992, Proceedings of 30th Liege International
Astrophysical Colloquium, eds Brahic A., Gerard J.C., Surdej J., 
Universit\'e de Li\`ege, Li\`ege, p. 285

\bibitem[Bauer et al.(2003)]{} Bauer J.~M., Meech K.~J., Fern\'andez
Y.~R, Pittichova~J., Hainaut O.~R., Boenhnhard H.\ 2003, Icarus,
166, 195

\bibitem[Boehnhardt et al.(2002)]{2002A&A...395..297B} Boehnhardt H.~et 
al. 2002, A\&A, 395, 297 

\bibitem[Chambers(1999)]{1999MNRAS.304..793C} Chambers J.~E.\ 1999, 
MNRAS, 304, 793 

\bibitem[Clube \& Napier(1984)]{1984MNRAS.211..953C} Clube S.~V.~M.,
Napier W.~M.\ 1984, MNRAS, 211, 953 

\bibitem[Dones, Levison, \& Duncan(1996)]{1996ciss.conf..233D} Dones L., 
Levison H.~F., Duncan M.\ 1996, ASP Conf.~Ser.~107: Completing the 
Inventory of the Solar System, 233 

\bibitem[Durda \& Stern(2000)]{2000Icar..145..220D} Durda D.~D., Stern
S.~A.\ 2000, Icarus, 145, 220

\bibitem[Emelyanenko(2004)]{emel}
Emel'yanenko V.V., Asher D.J., Bailey M.E. \ 2004, MNRAS, 350, 161

\bibitem[Evans \& Tabachnik(1999)]{1999Natur.399...41E} Evans N.~W.,
Tabachnik S.\ 1999, Nature, 399, 41 

\bibitem[Fern\'andez(1985)]{1985Icar...64..308F} Fern{\'a}ndez J.~A.\ 1985, 
Icarus, 64, 308 

\bibitem[Groussin et al.(2004)]{groussin}
Groussin O., Lamy P., Jorda L. 2004, A\&A, 413, 1163

\bibitem[Hahn \& Bailey(1990)]{1990Natur.348..132H} Hahn G.,
Bailey M.~E.\ 1990, Nature, 348, 132

\bibitem[Hainaut \& Delsanti(2002)]{2002A&A...389..641H} Hainaut O.~R., 
Delsanti A.~C.\ 2002, A\&A, 389, 641 

\bibitem[Holman (1997)]{matt}
Holman M.~J.\ 1997, Nature, 387, 785

\bibitem[Horner(2003)]{jonti}
Horner J., 2003, ``The Behaviour of Small Bodies in the Outer Solar
System'', D. Phil. thesis, University of Oxford.

\bibitem[Horner, Evans, Bailey, \& Asher(2003)]{2003MNRAS.343.1057H} 
Horner J., Evans N.~W., Bailey M.~E., Asher D.~J.\ 2003, MNRAS, 
343, 1057 

\bibitem[Horner, Evans \& Bailey(2004)]{2004MNRAS.343.new} 
Horner J., Evans N.~W., Bailey M.~E.\ 2004, MNRAS, submitted.

\bibitem[Lamy et al.(2004)]{lamy}
Lamy P.L., Toth I., Fern\'andez Y.R., Weaver H.A. 2004,
Comets II, M.~C.~Festou, U.~Keller, H.~A.~Weaver (eds), University of
Arizona Press, Tucson, in press

\bibitem[Levison \& Duncan(1994)]{1994Icar..108...18L} Levison H.~F.,
Duncan M.~J.\ 1994, Icarus, 108, 18 

\bibitem[Levison \& Duncan(1997)]{1997Icar..127...13L} Levison H.~F., 
Duncan M.~J.\ 1997, Icarus, 127, 13 

\bibitem[Lowry et al.(2003)]{LFC} Lowry S.~C., Fitzsimmons A.,
Collander-Brown S.\ 2003, A\&A, 397, 329

\bibitem[Morbidelli, Bottke, Froeschl{\' e}, \&
Michel(2002)]{2002aste.conf..409M} Morbidelli A., Bottke W.~F.,
Froeschl{\' e} C., Michel P.\ 2002, Asteroids III, W.~F.~Bottke Jr.,
A.~Cellino, P.~Paolicchi, and R.~P.~Binzel (eds), University of
Arizona Press, Tucson, p.409

\bibitem[Oikawa \& Everhart(1979)]{1979AJ} Oikawa S.,
Everhart E.\ 1979, AJ, 84, 134 

\bibitem[Opik (1976)]{Opik76}
Opik E.J. 1976 Interplanetary Encounters, Elsevier, New York

\bibitem[Peixinho et al.(2001)]{2001A&A...371..753P} Peixinho N.,
Lacerda P., Ortiz J.~L., Doressoundiram A., Roos-Serote M.,
Guti{\' e}rrez, P.~J.\ 2001, A\&A, 371, 753

\bibitem[Press et al]{press} Press W.H., Teukolsky S.A., Vetterling
W.T., Flannery B.P. 1992, Numerical Recipes (Cambridge University
Press: Cambridge), chap. 15

\bibitem[Sanzovo et al.(2001)]{2001MNRAS.326..852S} Sanzovo G.~C., de 
Almeida A.~A., Misra A., Torres R.~M., Boice D.~C., Huebner, W.~F.\ 
2001, MNRAS, 326, 852 

\bibitem[van Woerkom (1948)]{}
van Woerkom A.J. 1948 Bull. Astron. Inst. Netherlands, 10, 445

\bibitem[Weintraub, Tegler, \& Romanishin(1997)]{1997Icar..128..456W} 
Weintraub D.~A., Tegler S.~C., Romanishin W.\ 1997, Icarus, 128, 456 

\end{thebibliography}
\end{document}